\documentclass[3p,twocolumn,authoryear]{elsarticle}

\usepackage{amssymb}
\usepackage{amsmath}

\usepackage{multicol}
\usepackage{xcolor}
\journal{Journal of Theoretical Biology}

\begin{document}

\begin{frontmatter}

\title{Modeling tumor progression in heterogeneous microenvironments: A cellular automata approach} 

\author[label1]{Yue Deng}
\author[label1]{Mingjing Li}
\author[label2,label3]{Jinzhi Lei\corref{correspondingauthor}}
\affiliation[label1]{organization={School of Software},
             addressline={Tiangong University},
             city={Tianjin},
             postcode={300387},
             country={China}}
\affiliation[label2]{organization={School of Mathematical Sciences},
        	addressline={Tiangong University},
	        city={Tianjin},
         	postcode={300387},
	        country={China}}
\affiliation[label3]{organization={Center for Applied Mathematics},
	        addressline={Tiangong University},
         	city={Tianjin},
         	postcode={300387},
       	    country={China}}
\cortext[correspondingauthor]{Corresponding authors\\ Email address:  jzlei@tiangong.edu.cn (J. Lei)}

\begin{abstract}
Understanding how microenvironmental heterogeneity influences tumor progression is essential for advancing both cancer biology and therapeutic strategies. In this study, we develop a cellular automata (CA) model to simulate tumor growth under varying microenvironmental conditions and genetic mutation rates, addressing a gap in existing studies that rarely integrate these two factors to explain tumor dynamics. The model explicitly incorporates the cellular heterogeneity of stem and non-stem cells, dynamic cell-cell interactions, and tumor-microenvironment crosstalk. Using computational simulations, we examine the synergistic effects of gene mutation rate, initial tumor burden, and microenvironmental state on tumor progression. Our results demonstrate that lowering the mutation rate significantly mitigates tumor expansion and preserves microenvironmental integrity. Interestingly, the initial tumor burden has a limited impact, whereas the initial condition of the microenvironment critically shapes tumor dynamics. A supportive microenvironment promotes proliferation and spatial invasion, while inhibitory conditions suppress tumor growth. These findings highlight the key role of microenvironmental modulation in tumor evolution and provide computational insights that may inform more effective cancer therapies.
\end{abstract}


\end{frontmatter}

\section{Introduction}
\label{sec1}
Cancer progression is a complex, dynamic process shaped by interactions across multiple biological scales, from molecular alterations to tissue-level environmental changes \citep{Swanton:2024aa}. 
The clonal evolution theory, which posits that cancer develops through the accumulation of genetic and epigenetic mutations, has long served as a foundational framework in oncology \citep{Laplane:2024aa,Househam:2022aa}. However, recent high-throughput sequencing studies have revealed that cells carrying oncogenic mutations are frequently found in histologically normal tissues. \citep{Martincorena:2015aa,Ma:2024aa,Hashimoto:2024aa,Visser:2023aa}. These findings challenge the view that genetic mutations alone are sufficient to drive malignant transformation.

An increasing body of evidence suggests that the tumor microenvironment (TME) plays a pivotal role in regulating cancer development and progression \citep{Beyond24Shao}. For instance, tissue abnormalities initiated by oncogenic mutations can be reversed by surrounding normal tissue \citep{Brown:2017aa,Moya:2019aa,Hill:2021aa}. Similarly, patients with chronic myeloid leukemia (CML) can remain in long-term remission despite persistent leukemic stem cells, a phenomenon known as treatment-free remission (TFR) \citep{Bourne:2024aa,Mahon:2010aa,Saussele:2018aa,Etienne:2017aa,Shah:2020aa}. These observations highlight the crucial role of the microenvironment in modulating the fate of potentially malignant cells. Therefore, understanding how microenvironmental heterogeneity influences tumor evolution is essential for developing effective prevention and therapeutic strategies.

The TME and tumor cells together form a dynamic ecosystem, wherein their interactions can either suppress or promote tumor progression. Quantitative modeling has become a vital tool in elucidating these complex dynamics. A variety of mathematical models have been proposed to study the influence of the TME on tumor growth and morphology, incorporating mechanisms such as cell-cell and cell-matrix interactions \citep{Anderson:2005aa,Liang:2019aa}, cell competition  \citep{Frieboes:2006aa,Zhang:2022aa}, spatial variation in environmental conditions \citep{Anderson:2006aa,Lai:2024aa}, and tumor-immune system interactions \citep{Li:2025hd}. These models offer mechanistic insights into the interplay between genetic and environmental factors in tumor development.

While continuum-based models, including those based on ordinary and partial differential equations (ODEs and PDEs), have contributed significantly to our understanding of therapy optimization and population-level kinetics \citep{Savageau1980GrowthEA,Sachs2001SimpleOM,Pillis2013ACA,Benzekry2014ClassicalMM,Hartung2014MathematicalMO,Polovinkina2021StabilityOS}, they inherently rely on mean-field approximations that smooth out local spatial irregularities and stochastic events. However, tumor initiation is often driven by rare events in specific local niches, which are difficult to capture in homogenized continuum frameworks. In contrast, cellular automata (CA) models represent cells as discrete agents on a lattice, offering distinct advantages for the objective of this study. Specifically, the CA framework allows us to: (1) biologically grounded individual cell behaviors (e.g., stochastic mutation and state transitions); (2) explicitly represent the spatial heterogeneity of the tissue; and (3) simulate the localized, bidirectional interactions between cells and their immediate microenvironment. These characteristics are particularly suited for exploring how local rules give rise to emergent tumor-level phenomena \citep{Deutsch2020BIOLGCAAC,Weerasinghe2019MathematicalMO,Metzcar2019ARO,Jamali2010ASV}.

Despite progress in modeling tumor heterogeneity, the specific role of the microenvironmental deterioration kinetics in driving the transition from normal to malignant phenotypes remains underexplored. In this study, we present a stochastic CA-based computational model to investigate tumor progression. The primary goal of this model is to provide qualitative mechanistic insights into tumor-microenvironment interactions, rather than to serve as a quantitative predictive tool for specific clinical cases. To this end, the specific objectives of this study are summarized as follows:
\begin{enumerate}
\item \textbf{Methodological Construction:} To develop a discrete model that integrates cancer cell heterogeneity (stem vs. non-stem), plasticity, and dynamic feedback loops with the surrounding microenvironment.
\item \textbf{Biological Investigation:} To systematically examine how the synergy between genetic mutation rates and initial environmental conditions determines tumor fate.
\item \textbf{Mechanistic Insight:}  To demonstrate that the TME acts as a critical checkpoint, where specific kinetic parameters of environmental deterioration can either suppress oncogenesis or catalyze malignant expansion.
\end{enumerate} 
\noindent Our results demonstrate that, alongside genetic mutations, the microenvironment's state critically influences tumor dynamics. These findings underscore the importance of targeting microenvironmental factors in cancer prevention and therapy, and provide mechanistic insights into the coupled roles of intrinsic and extrinsic influences in tumor evolution. 

\section{Model and Methods}
\label{sec2}
We developed a discrete computational model to simulate tumor evolution originating from normal tissue cells in a localized region of a living organism. The model captures both the spatial structure of tissue and the state dynamics of individual cells, comprising two core components: (1) a two-dimensional cellular automaton (CA) that defines spatial relationships and cell positioning, and (2) a set of rules governing state transitions and interactions among different cell types, incorporating the effects of a dynamic microenvironment.

This section describes the model structure, transition rules, and numerical implementation in detail.

\subsection{Cellular automaton model}
\label{subsec1}
Our model employs a two-dimensional cellular automaton (CA) to simulate the spatiotemporal dynamics of tumor progression. The CA is defined on a hexagonal grid, where each site represents a local spatial position with six immediate neighbors. Each site can either be empty or occupied by a single cell. A cell may adopt one of four states: normal stem cell ($N_1$), normal cell ($N_2$), tumor stem cell ($T_1$), and tumor cell ($T_2$).

\begin{figure}[htbp]
	\centering
	\includegraphics[width=8cm]{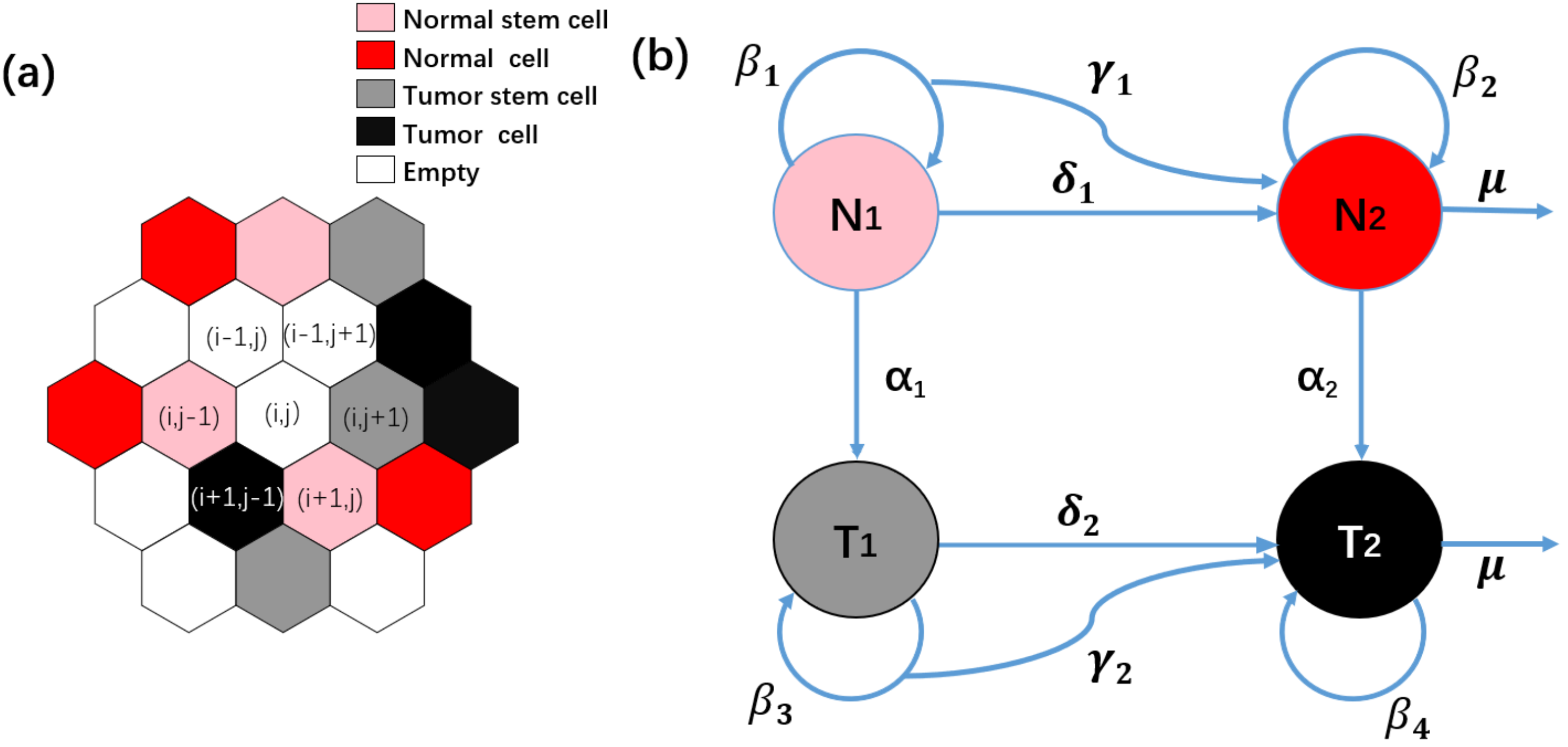}
	\caption{\textbf{Spatial structure of the cellular automaton and cell-type transition diagram.} \textbf{(a)} Schematic of the two-dimensional cellular automaton, where each hexagonal grid cell can be in one of five states: empty, normal stem cells ($N_1$), normal cell ($N_2$), tumor stem cell ($T_1$), or tumor cell ($T_2$). \textbf{(b)} Diagram of cell-type transitions. All cell types are capable of self-renewal, while stem cells ($N_{1}$, $T_1$) can also differentiate into non-stem cells ($N_{2}$, $T_2$). Non-stem cells can undergo death at a rate $\mu$. Normal cells may acquire mutations and convert to tumor cells, with $\alpha_{1}$ and $\alpha_{2}$ denoting mutation rates for normal stem cells and normal non-stem cells, respectively. The heterogeneous microenvironment is represented by a continuous variable $m$ ($0 < m< 1$) assigned to each grid point. The microenvironment is dynamically influenced by the local densities of tumor and normal cells, and all cell-state transition rates are modulated by the local microenvironmental index.}
	\label{fig:1}
\end{figure}

Figure \ref{fig:1}a illustrates the CA structure and the associated cell types. To simulate tumor progression, the grid is initially populated with a specific cell distribution. Cellular processes--such as growth, death, and mutation--drive the temporal evolution of the cell population.

The CA model evolves over discrete time steps to capture the transition from healthy tissue to malignancy. At each step, every cell undergoes one of four possible behaviors: interphase (quiescent), division, differentiation, or death. Cell fate decisions are governed by both intrinsic properties (e.g., stemness, proliferative potential) and extrinsic factors (e.g., neighboring cell density, local environmental conditions). By iteratively updating these behaviors, the model simulates the spatiotemporal dynamics of tumor initiation and expansion.

Figure \ref{fig:1}b depicts the state transition diagram for the four cell types. All cell types are capable of self-renewal, while non-stem cells are also subject to death. Stem cells can differentiate into non-stem cells. Additionally, mutations may occur in both stem and non-stem normal cells, driving transitions to tumor phenotypes. These stochastic transitions are governed by mutation rates $\alpha_1$ and $\alpha_2$ for stem and non-stem cells, respectively.

To capture spatial heterogeneity, we introduce a microenvironmental index $m$ at each grid location. This index evolves continuously based on the local composition. In turn, the microenvironment modulates cell behavior by influencing transition rates, including proliferation, differentiation, and cell death. This feedback loop allows the model to reflect the co-evolution of the tumor and its tissue microenvironment.

\subsection{Methods}
\subsubsection{Cell behaviors}

To simulate tissue growth, the model incorporates key cellular behaviors: proliferation, differentiation, and death. Stem cells cycle between two phases: interphase and division. Differentiation can occur in either phase, though the rate is lower during interphase compared to the division phase.

\vspace{0.25cm}
\textit{Proliferation and differentiation.}\ \  During the cell cycle, stem cells in interphase can either differentiate into non-stem cells (normal stem cells at a rate $\delta_{1}$; tumor stem cells at a rate $\delta_{2}$), or proceed to the division phase at a rate $\beta$. In the division phase, a stem cell divides into two daughter cells. Following division, each daughter cell may differentiate into a non-stem cell (at rate $\gamma_1$ or $\gamma_2$), remain a stem cell, or undergo mutation and return to interphase. Spatially, one daughter cell effectively occupies the original grid location, while the other is placed in a neighboring site. If multiple neighboring sites are empty, one is randomly selected; if no space is available, cell division is suppressed (contact inhibition).

Cell proliferation is regulated by local cell density and microenvironmental conditions. Neighboring cells may secrete inhibitory cytokines that suppress proliferation. Consequently, the effective cell proliferation rate $\beta$ decreases with increasing local cell density. This inhibitory effect is modeled using a Hill function:
\begin{equation}
	\label{eq:1}
	\beta = \beta_{0}\times\frac{1}{1+(N_{q}/\theta)^{s_{0}}},
\end{equation}
where $\beta_0$ is the maximum proliferation rate, $N_q$ is the total number of occupied neighboring sites (local density), $\theta$ is a saturation constant modulated by the microenvironment, and $s_0$ is the Hill coefficient.

The saturation level $\theta$ depends on the local microenvironment index $m$ (with $0 < m < 1$), which quantifies the favorability of the environment for tumor development (higher $m$ indicates a pro-tumor environment). The dependence of $\theta$ on $m$ differs between normal and tumor cells:
\begin{equation}\label{eq:2}
	\theta = \begin{cases}
		\theta_{0}+\theta_{1}\frac{\theta_{2}^{s_{1}}}{\theta_{2}^{s_{1}}+ m^{s_{1}}}, &  \mbox{for normal cells}\\
		\theta_{0}+\theta_{1}\frac{m^{s_{1}}}{\theta_{2}^{s_{1}}+ m^{s_{1}}}, &  \mbox{for tumor cells},
	\end{cases}
\end{equation}
where $\theta_{0}$, $\theta_{1}$, $\theta_{2}$, and $s_{1}$ are model parameters. For normal cells, increasing $m$ reduces $\theta$, thereby suppressing proliferation. Conversely, for tumor cells, increasing $m$ raises $\theta$, promoting cell division.

\vspace{0.25cm}
\textit{Cell death and fitness.}\ \ To model non-stem cell death, we introduce a fitness function $g(m)$ reflecting adaptability to the local microenvironment. Tumor cells thrive in high-$m$ environments, while normal cells favor low-$m$ conditions. Fitness is modeled linearly as:
\begin{equation}
	\label{eq:3}
	g(m) = \begin{cases}
		a (1-m), & \mbox{for normal cells}\\
		a m, & \mbox{for tumor cells},
	\end{cases}
\end{equation}
where $a$ is a sensitivity constant.

The cell death rate $\mu$ is inversely related to fitness so that higher fitness implies lower death rates, which is represented via a logistic function:
\begin{equation}
	\label{eq:4}
	\mu= \frac{\mu_{0}}{1+ c\times e^{g}},
\end{equation}
where $\mu_{0}$ is the maximum death rate (cell-type specific) and $c$ is a scaling constant. Higher fitness increases the denominator, thereby reducing the overall death rate.

\subsubsection{Evolution of the microenvironment}

The tumor microenvironment (TME) is highly heterogeneous and evolves dynamically through interactions with resident cells. Tumor cells actively remodel their niche to support survival and proliferation, often by suppressing immune response and altering signaling landscapes \citep{Wellenstein:2018aa,Peng:2015aa,Suvac:2025aa}.

To model this, we utilize the continuous variable $m$ ($0<m<1$) at each grid point. Higher $m$ values represent pro-tumor conditions, while lower values represent anti-tumor (normal) conditions. The local evolution of $m$ is driven by the density of surrounding tumor and normal cells: tumor cells drive $m$ upward (pro-tumor), while normal cells drive $m$ downward (anti-tumor). The dynamic equation is given by
\begin{equation}
	\label{eq:5}
	\frac{d m}{d t}= k_{1}(1-m)-k_{2}m,
\end{equation}
where $k_1$ is the tumor-induced transition rate, and $k_2$ is the normal-cell-induced transition rate. These rates depend on the local proportions of tumor ($R_c$) and normal ($R_n$) cells, defined as:
\begin{equation}
	\left\{
	\begin{aligned}
		k_{1}(R_{c})&= k_{11}\frac{R_{c}^{n_{1}}}{k_{12}^{n_{1}}+ R_{c}^{n_{1}}},\\
		k_{2}(R_{n})&= k_{21}\frac{R_{n}^{n_{2}}}{k_{22}^{n_{2}}+ R_{n}^{n_{2}}},
	\end{aligned}
	\right.
\end{equation}
where $k_{11}$, $k_{12}$, $k_{21}$, $k_{22}$, $n_1$, and $n_2$ control the sensitivity and saturation of the environmental response to cell densities.

\subsubsection{Numerical scheme}
\label{sec:2.2.3}

The equations defined above govern the rules for cell division, differentiation, death, mutation, and microenvironmental evolution. To simulate the system dynamics, we initialize the cellular automaton by randomly distributing specific cell types across the grid and assigning an initial microenvironmental value $m$ sampled uniformly from a specified range for each grid point.

The simulation proceeds in discrete time steps $\Delta t$. At each step, a stochastic update process is applied to determine cell fate. Specifically, for each cell, a random number $r$ is generated using the standard uniform random number generator (e.g., \texttt{rand( )} in \texttt{Python}) and compared with the cumulative probabilities of possible behaviors. Based on this comparison, the cell fate is determined. The following update rules applied to each cell and grid location are as follows:
\begin{enumerate}
	\item \textbf{Differentiation:} Stem cells in the interphase may differentiate into non-stem cells with probability $\delta \Delta t$, where $\delta$ depends on the cell phenotype.
	\item \textbf{Division:} Each cell undergoes division with probability $\beta \Delta t$, where $\beta$ is computed from Eq.~\eqref{eq:1}. For dividing stem cells, each daughter cell may differentiate into a non-stem cell with probability $\gamma \Delta t$, or remain a stem cell and return to the interphase.
	\item \textbf{Cell death:} Non-stem cells may die with probability $\mu \Delta t$, where $\mu$ is defined by Eq. ~\eqref{eq:4}.
	\item \textbf{Mutation:} Normal cells may undergo mutation with probability $\alpha_1\Delta t$ (for stem cells) or $\alpha_2 \Delta t$ (for non-stem cells), driving a transition to tumor phenotypes.
	\item \textbf{Microevironment update:} The local microenvironmental value $m$ is updated according to Eq.~\eqref{eq:5}, reflecting the influence of surrounding cell populations.
\end{enumerate}

This stochastic simulation framework captures the coupled temporal evolution of cell populations and their microenvironment. The overall flowchart of the stochastic process is illustrated in Figure \ref{fig:2}, with code implementation details in Appendix A. The model allows for a wide range of biological scenarios by adjusting mutation rates, initial conditions, and interaction parameters. In particular, it enables the exploration of spatial competition between normal and tumor cells under different microenvironmental conditions.

All simulations were performed on a two-dimensional grid of size $100\times 100$, representing a tissue region with up to $10^4$ cells. The simulation time step was set to $\Delta t = 1\mathrm{h}$. 

\begin{figure*}[htbp]
	\centering
	\includegraphics[width=13cm]{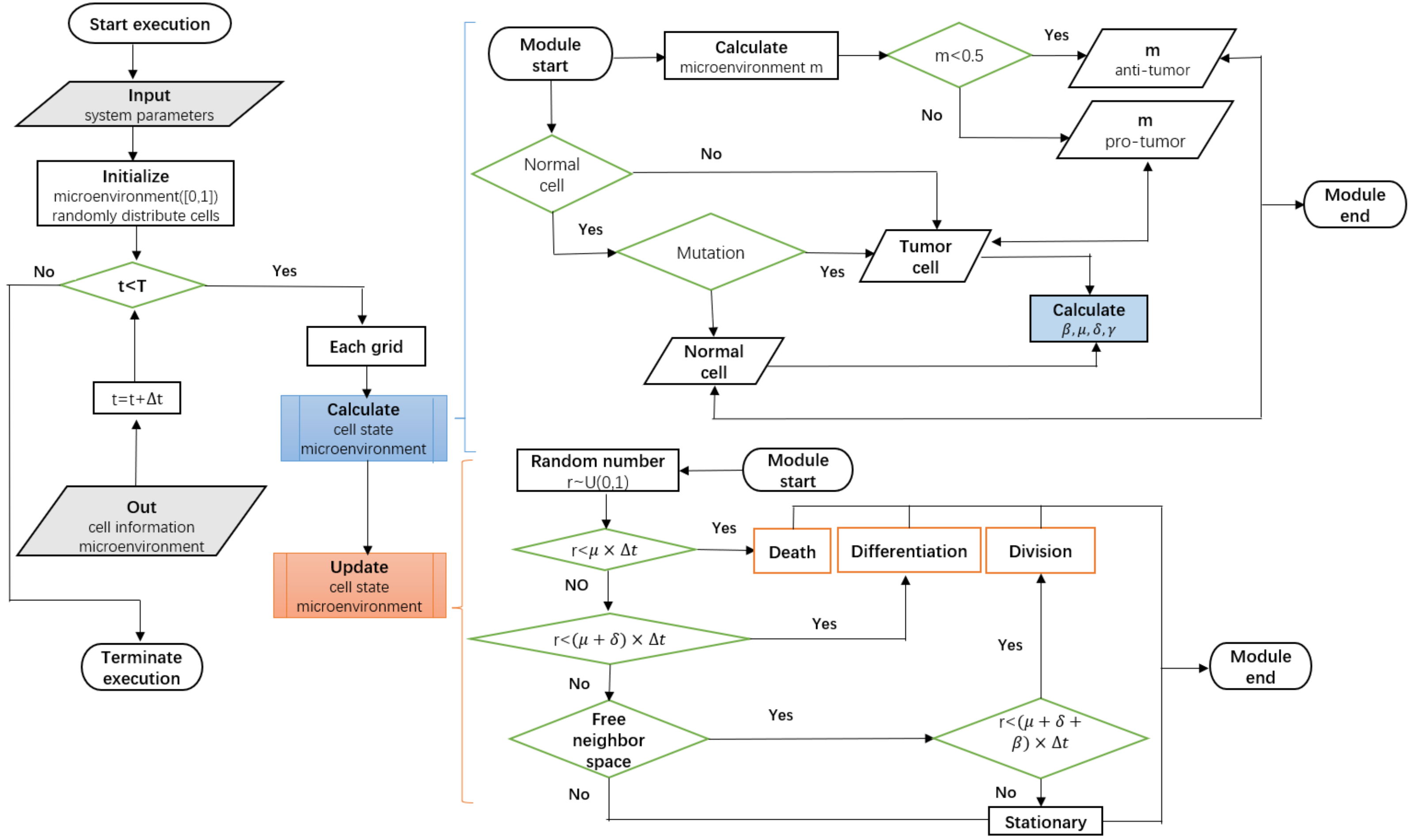}
	\caption{\textbf{Overview of the stochastic simulation algorithm.} The left panel depicts the main flowchart of the simulation loop. The right panels detail the two core sub-modules: the calculation of update rates for cell states and the microenvironment, and the execution of state updates for current cells.}
	\label{fig:2}
\end{figure*}

Given the stochastic nature of the model, we assessed the reproducibility of the simulation outcomes. We observed that the long-term evolutionary trends are highly consistent across independent realizations. Consequently, single representative runs are utilized to visualize the characteristic temporal dynamics and spatial patterns. For quantitative comparisons and to assess variability, aggregate results are reported as the mean $\pm$ standard deviation derived from $n = 5$ independent simulation runs.

\subsubsection{Parameter estimation}
\begin{figure}[htbp]
	\centering
	\includegraphics[width=7cm]{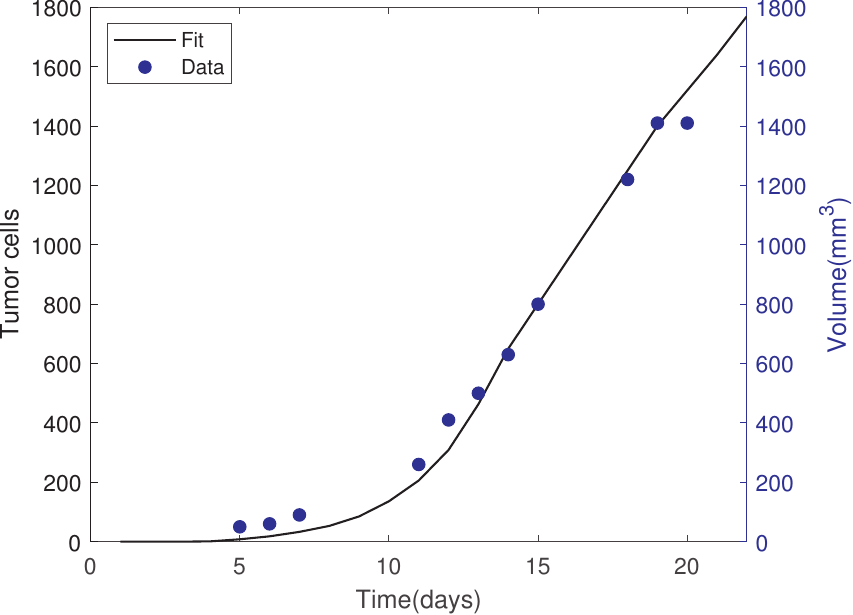}
	\caption{\textbf{Calibration of model parameters against experimental data.} Data on tumor volume derived from lung cancer \citep{Benzekry2014ClassicalMM} are shown as blue dots, and the corresponding model simulation results are shown as a black solid line.}
	\label{fig:3}
\end{figure}

To ensure biological relevance, model parameters were estimated based on literature values and experimental data related to tumor growth and cell behavior. The parameters were adjusted to reproduce key features observed in experimental tumor development. Specifically, we calibrated the model using longitudinal tumor volume data from lung cancer studies \citep{Benzekry2014ClassicalMM}. Figure \ref{fig:3} illustrates the concordance between the temporal dynamics of the simulated tumor cell population and the experimental data.

Initial parameter values were selected by referring to previous studies. Parameters were then refined by comparing simulation results with the experimental dynamics of tumor growth. The time scale of tumor growth and the ratio of various cell types were used as key criteria for parameter estimations. Parameter estimations were performed in two stages:
\begin{enumerate}
	\item \textbf{Baseline estimation:} Parameters governing normal cells and microenvironment evolution were calibrated using data from non-mutated cell populations.
	\item \textbf{Tumor-specific estimation:} Additional parameters related to mutation, tumor proliferation, and microenvironmental interaction were fitted using tumor growth data.
\end{enumerate}

Table~\ref{tab:1} summarizes the full set of model parameters used in the simulations.

\begin{table*}[t]
	\centering
	\caption{Model parameters of the cellular automaton and cell state transitions. All time-related parameters are given in hours (h).}
	\label{tab:1}
	\begin{tabular}{p{1.3cm}p{9cm}p{1.0cm}p{2.8cm}}
		\hline
		Parameter  &  Description    &  Value   & Source  \\ 
		\hline
	$m$ &  Microenvironmental index at each grid location & $0\sim1$ & Regulation\\
		$\beta_{1}$  &  Maximum proliferation rate of normal stem cells &  0.034 & Estimated\\
		$\beta_{2}$  &  Maximum proliferation rate of normal cells &  0.105  & \citet{Zhang2022Optimal}$^{(a)}$\\
		$\beta_{3}$  &  Maximum proliferation rate of tumor stem cells &  0.035 & Estimated \\
		$\beta_{4}$  &  Maximum proliferation rate of tumor cells &  0.192  & \citet{Jiao2011EmergentBF}$^{(a)}$\\
		$\delta_{1}$  &  Differentiation rate of normal stem cells during the interphase &  0.003 &  Estimated\\
		$\delta_{2}$  &  Differentiation rate of tumor stem cells during the interphase&  0.0025 &  Estimated\\
		$\gamma_{1}$  &  Differentiation rate of normal stem cells during the division phase &  0.007 & Estimated \\
		$\gamma_{2}$  &  Differentiation rate of tumor stem cells during the division phase &  0.0065 & Estimated \\
		$\alpha_{1}$  &  Mutation rate of normal stem cells &  0.00022  &  Estimated\\
		$\alpha_{2}$  &  Mutation rate of normal cells &  0.0004  &  Estimated\\
		$D_{1}$  &  Maximum death rate of normal cells &  0.05  &  Estimated\\
		$D_{2}$  &  Maximum death rate of tumor cells &  0.062  &  Estimated\\
		$s_{0}$  &  Hill coefficient in the proliferation rate &  1 & \citet{Zhang2022Entropy}\\
		$s_{1}$  &  Hill coefficient in the proliferation rate &  8.5 & \citet{Zhang2022Entropy}\\
		$\theta_{0}$  &  The coefficient in $\theta(m)$ &  300  & \citet{Zhang2022Entropy}\\
		$\theta_{1}$  &  The coefficient in $\theta(m)$ &  1000 & \citet{Zhang2022Entropy} \\
		$\theta_{2}$  &  The coefficient in $\theta(m)$ &  0.4  & \citet{Zhang2022Entropy}\\
		$a$  &  The coefficient of the fitness function &  0.8  & \citet{Zhang2022Entropy}\\
		$c$  &  The coefficient of the death rate of cells  &  1  & \citet{Zhang2022Entropy}\\
		$k_{11}$  &  The coefficient of microenvironment transition rate of normal stem cells and normal cells &  1  &  Estimated \\
		$k_{12}$  &  The coefficient of microenvironment transition rate of normal stem cells and normal cells &  5   &  Estimated\\
		$n_{1}$  &  The coefficient of microenvironment transition rate of normal stem cells and normal cells &  3.2  &  Estimated \\
		$k_{21}$  &  The coefficient of microenvironment transition rate of tumor stem cells and tumor cells &  1   &  Estimated\\
		$k_{22}$  &  The coefficient of microenvironment transition rate of tumor stem cells and tumor cells &  5   &  Estimated\\
		$n_{2}$  &  The coefficient of microenvironment transition rate of tumor stem cells and tumor cells &  3.2   &  Estimated\\
		\hline
	\end{tabular}
\begin{minipage}{14cm}
	\begin{enumerate}
		\item[$^{(a)}$] Note: The proliferation rate of tumor cells ($\beta_{4}$) is set larger than that of normal cells ($\beta_{2}$) to reflect the enhanced proliferative capacity characteristic of malignant phenotypes.
	\end{enumerate}
\end{minipage}
\end{table*}

\subsection{Model Validation}
\label{sec:validation}
To validate the biological plausibility of our model, we compared simulation outcomes with general principles of tumor growth. First, we verified that the model maintains normal tissue homeostasis when the mutation rate is set to zero, consistent with the stability of healthy tissue. Second, we observed that the tumor cell population size is positively correlated with the mutation rate. Critically, simulation results indicate that long-term tumor evolution dynamics are independent of the initial cell count, provided the simulation duration is sufficient. Finally, the expansion of the tumor population exhibits a strong dependence on microenvironmental conditions: favorable environments (low $m$) inhibit tumor growth, whereas adverse environments (high $m$) promote malignant expansion. 

\section{Results}
\subsection{Evolution dynamics in the absence of gene mutation}
\label{sec:nomut}

To evaluate the performance of the proposed model, we first investigated tissue dynamics in the absence of gene mutations by setting the mutation rate to zero. We began by simulating the growth of normal tissue, initializing the system with $20$ normal stem cells and a spatially heterogeneous microenvironment in which each grid was randomly assigned a value $m\in(0, 0.3)$.

\begin{figure*}[htbp]
	\centering
	\begin{center}
		\includegraphics[width=13cm]{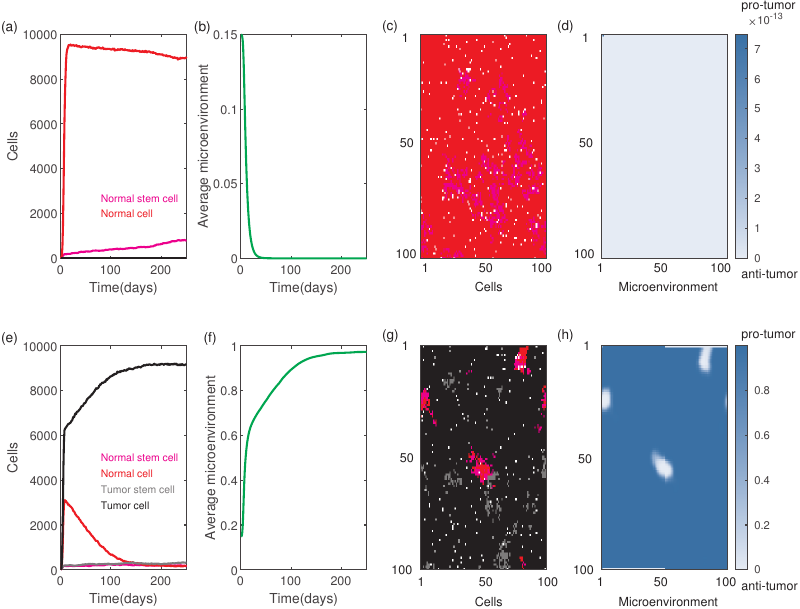}
	\end{center}
	\caption{\textbf{Evolution dynamics of cells and the microenvironment in the absence of gene mutations}. \textbf{Top row (a-d):} Simulation initialized with $20$ normal stem cells and a randomly distributed microenvironment $m \in (0, 0.3)$. \textbf{Bottom row (e-h):} Simulation initialized with $20$ normal stem cells and $20$ tumor stem cells, with the same initial microenvironment range. \textbf{(a, e)} Temporal evolution of cell populations. \textbf{(b, f)} Temporal evolution of the average microenvironment value. \textbf{(c, g)} Spatial distribution of all cell types at day $250$. \textbf{(d, h)} Spatial distribution of the microenvironment at day $250$. Results represent a single stochastic realization.}
	\label{fig:4}
\end{figure*}

Figure \ref{fig:4}a shows the time course of the cell populations. Due to their lower proliferation rate, normal stem cells expanded slowly, reaching $807$ cells by day $250$. In contrast, normal cells proliferated rapidly, peaking on day $20$, and subsequently stabilized at approximately $8,937$ cells, representing $86\%$ of the total population. The spatial distribution of cells at day $250$ (Figure \ref{fig:4}c) confirms that normal cells (red) dominate the tissue, with a small fraction of normal stem cells (pink) scattered throughout the grid.

The evolution of the microenvironment (Figure \ref{fig:4}b) mirrors these cellular dynamics. As the cell population expands, the average microenvironmental value declines and eventually stabilizes at a low level. This transition reflects the influence of normal cells, which drive the local environment toward an anti-tumor state. By day $250$, the spatial distribution of the microenvironment (Figure \ref{fig:4}d) becomes more homogeneous, indicating uniformly favorable conditions that support the maintenance of normal tissue homeostasis.

Next, we explored the effect of introducing the tumor cells by initializing the system with both $20$ normal stem cells and $20$ tumor stem cells, while maintaining the same randomly distributed microenvironment ($m \in (0, 0.3)$). The resulting dynamics are presented in Figure \ref{fig:4}e-h. In the early stages, both normal and tumor cells expanded rapidly. However, as the simulation progressed, the normal cell population sharply declined while tumor cells continued to proliferate. By day $250$, tumor cells constituted more than $92\%$ of the total population, whereas both normal and tumor stem cells remained a minority due to their relatively low proliferation rates.

The spatial distribution at day $250$ (Figure \ref{fig:4}g) clearly shows tumor cells occupying the majority of the grid, with only a few isolated pockets of normal cells (stem or non-stem) remaining. This pattern highlights the competitive advantage of tumor cells in securing space and resources.

Correspondingly, the microenvironment underwent significant changes. As tumor cells expanded, the average value of $m$ increased steadily, reaching $0.97$ by day $250$ (Figure \ref{fig:4}f). The spatial map of the microenvironment (Figure \ref{fig:4}h) indicates a nearly uniform shift toward pro-tumor conditions, particularly in regions heavily populated by tumor cells. A comparison between Figures \ref{fig:4}g and \ref{fig:4}h reveals a strong spatial correlation between tumor cell dominance and elevated microenvironmental values, demonstrating how tumor cells actively reprogram their surroundings to favor continued growth and suppression of normal cells.

Collectively, these results indicate that even in the absence of \textit{de novo} gene mutations, the initial seeding of a tumor stem cell subpopulation is sufficient to trigger malignant expansion, displace normal cell populations, and remodel the microenvironment to favor tumor progression.

\subsection{Evolution dynamics with gene mutation}
	
We next investigated how gene mutations influence cell population dynamics by varying the mutation rates $(\alpha_1, \alpha_2)$ for normal stem cells and normal non-stem cells, respectively. Consistent with biological evidence that downstream progenitors accumulate more mutations than stem cells \citep{2007Identification}, we assumed $\alpha_2 > \alpha_1$. Simulations were initialized with $20$ normal stem cells and a randomly distributed microenvironment $m \in (0, 0.3)$, spanning a duration of $330$ days.
	
\begin{figure*}[htbp]
	\centering
	\includegraphics[width=13cm]{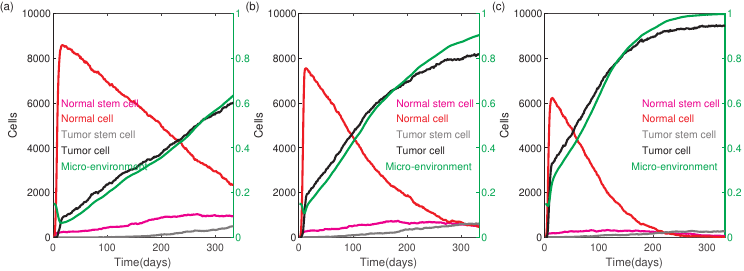}
	\caption{\textbf{Evolution dynamics of cells and microenvironment under varying gene mutation rates.} Mutation rates for normal stem cells ($\alpha_1$) and normal non-stem cells ($\alpha_2$) were set as follows:  \textbf{(a)} $(\alpha_1, \alpha_2) = (1.1\times 10^{-4}, 2.0\times 10^{-4})$. \textbf{(b)} $(\alpha_1, \alpha_2) = (2.2\times 10^{-4}, 4.0\times 10^{-4})$. \textbf{(c)} $(\alpha_1, \alpha_2) = (4.4\times 10^{-4}, 8.0\times 10^{-4})$. Data are shown from a representative single simulation for each mutation rate combination.}
	\label{fig:5}
\end{figure*}

Figure \ref{fig:5} illustrates the temporal trajectories of cell populations and microenvironmental conditions under different mutation regimes. Initially, the normal cell population expanded rapidly, but subsequently declined as tumor cells emerged and proliferated. Concurrently, the average microenvironmental value initially decreased---reflecting the homeostatic influence of normal tissue growth---but later reversed course, rising steadily as tumor cells dominated the system, eventually establishing a strongly pro-tumor environment.
	
As the mutation rates increased from $(\alpha_1, \alpha_2) = (1.1\times 10^{-4}, 2.0\times 10^{-4})$ to $(4.4\times 10^{-4}, 8.0\times 10^{-4})$, the tumor burden increased substantially, while the normal cell population was progressively suppressed. Specifically, the peak abundance of normal cells dropped from $8,590$ to $6,233$, and the timing of this peak shifted earlier, from day $18$ to day $12$. Higher mutation rates also accelerated the deterioration of the microenvironment. These results suggest that sustained genetic instability not only hastens tumor onset but also exacerbates environmental remodeling, facilitating rapid malignant dominance.

\begin{figure*}[t]
	\centering
	\includegraphics[width=13cm]{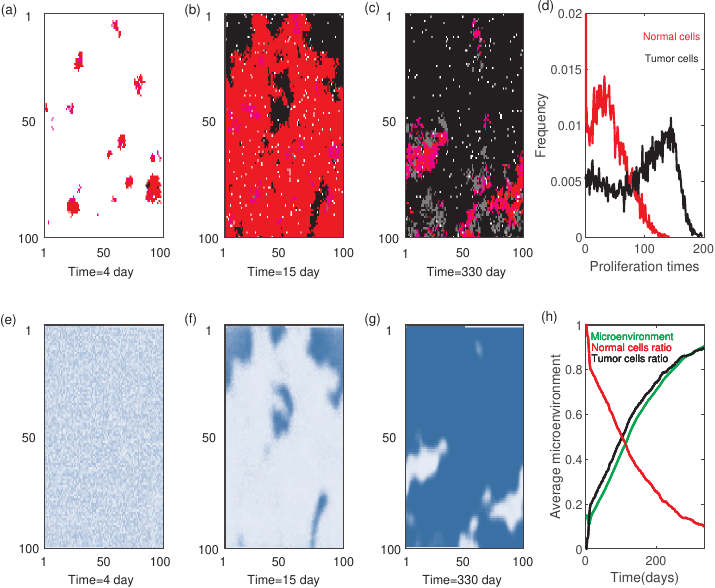}
	\caption{\textbf{Temporal and spatial evolution of cell populations and the microenvironment.} Spatial distribution of cells at: \textbf{(a)} day $4$, \textbf{(b)} day $15$, \textbf{(c)} day $330$. \textbf{(d)} Frequency distribution of cell proliferation cycles. Spatial distribution of the microenvironment at: \textbf{(e)} day $4$, \textbf{(f)} day $15$, \textbf{(g)} day $330$. \textbf{(h)} Temporal evolution of the average microenvironment value. Results represent a single stochastic realization. See Supplementary Video 1 in the online supplementary material for the dynamic evolution of cell distributions.}
	\label{fig:6}
\end{figure*}

To dissect the spatiotemporal evolution of tumor development, we further analyzed the intermediate case $(\alpha_1, \alpha_2) = (2.2\times 10^{-4}, 4.0\times 10^{-4})$. Based on the normal cell population trajectory shown in Figure \ref{fig:5}b, we selected three representative time points: days $4$ (early phase), $15$ (transition phase), and $330$ (late phase). Figures \ref{fig:6}a-c display the corresponding spatial distributions. On day $4$, the tissue was primarily composed of normal cells, with only sporadic tumor cells appearing. By day $15$, both populations had expanded, yet normal cells remained dominant. However, by day $330$, tumor cells---leveraging their higher proliferation rates---had outcompeted normal cells and colonized the majority of the spatial domain. These results demonstrate that even when originating from rare stochastic mutation events, tumor cells can eventually dominate the system through sustained proliferative advantage.

To quantify this replicative potential, we tracked the number of division cycles undergone by progeny of each stem cell lineage. Figure \ref{fig:6}d presents the frequency distribution of these cycle counts. Tumor cells exhibited a significantly broader distribution with higher peak cycle counts compared to normal cells, indicating both enhanced proliferative capacity and reduced apoptotic clearance. This survival advantage allows tumor cells to displace normal populations over time.

Figures \ref{fig:6}e-g depict the co-evolution of the microenvironment at the corresponding time points. As tumor cells expanded, the local microenvironment progressively deteriorated, with elevated $m$ values spatially coinciding with tumor-rich regions. In contrast, areas retaining normal cell dominance generally preserve more favorable (lower $m$) conditions. Figure \ref{fig:6}h summarizes the temporal trend: early in the simulation, when normal cells predominate, the microenvironment remains relatively homeostatic; however, as tumor cells expand, they drive the system toward a pro-tumor state. These findings underscore the dynamic interplay between cell population kinetics and environmental plasticity, emphasizing that gene mutations act as the initial trigger that sets in motion a self-reinforcing cycle of tumor expansion and microenvironmental remodeling.

\subsection{Influence of initial conditions on cell dynamics}
\label{sec:ini}

To assess the extent to which tumor progression depends on initial cellular and environmental configurations, we conducted simulations under four distinct scenarios. These scenarios varied by cell type composition and the initial state of the microenvironment (developing vs. homeostatic), allowing us to quantify the suppressive capacity of normal tissue. Specifically:
\begin{itemize}
	\item \textbf{Case 1 (Co-culture in developing tissue):} The simulation was initialized with a mixture of $20$ normal stem cells and $20$ tumor stem cells in a sparse environment. The microenvironment $m$ was randomly distributed in $(0, 0.3)$, and default mutation rates were applied $(\alpha_1, \alpha_2) = (2.2\times 10^{-4}, 4.0\times 10^{-4})$.
	\item \textbf{Case 2 (Mutation in developing tissue):} The system began with a sparse population of only $20$ normal stem cells (simulating tissue growth). The microenvironment $m \in (0, 0.3)$ and the default mutation rates were applied.
	\item \textbf{Case 3 (Mutation in mature tissue):} The initial condition was derived from the stable homeostatic state (final state of Figure \ref{fig:4}a-d), representing a healthy, space-filled tissue. Default mutation rates were applied to simulate spontaneous tumorigenesis.
	\item \textbf{Case 4 (Seeding in mature tissue without mutation):} The initial condition was the same homeostatic state as Case 3, into which $20$ tumor stem cells were artificially seeded. Crucially, gene mutation rates were set to zero to isolate the inhibitory effect of the healthy microenvironment on pre-existing tumor seeds.
\end{itemize}

Simulations for each case were run until tumor cells occupied at least $80\%$ of the grid. Table \ref{tab:2} summarizes the final population proportions and the time required to reach this tumor-dominant state. While tumor cells eventually dominated in all scenarios, the kinetics of progression varied drastically.

Case 1 displayed the fastest kinetics, reaching dominance in only $1$ month (Table \ref{tab:2}). This rapid expansion is driven by the immediate presence of tumor stem cells, nonzero mutation rates, ample free space, and loose microenvironmental structure, establishing a baseline for maximum tumor growth potential.

Case 2 and Case 3 illustrate the delay caused by the need for \textit{de novo} mutations and the impact of tissue density. In Case 2 (developing tissue), tumor onset took $14$ months. However, in Case 3 (mature tissue), the time to dominance extended to $26$ months significantly. This suggests that a mature, homeostatic tissue architecture acts as a physical and environmental barrier, delaying the expansion of mutant clones compared to a developing tissue.

Most notably, Case 4 demonstrates the profound suppressive role of the healthy microenvironment. When tumor stem cells were introduced into a stable healthy tissue in the absence of further mutations, it took $63$ months for the tumor to dominate---more than nearly five years of simulation time. This duration is over four times longer than Case 2 and more than double that of Case 3. This result indicates that even when tumor seeds are present, a robust healthy microenvironment (characterized by low $m$ and established normal cell interactions) can impose a powerful check on malignant growth, significantly prolonging the latency period before overt tumor formation.

\begin{table*}[htbp]
	\centering
	\caption{Cell population proportions at the final time point and simulation duration for different initial conditions}
	\begin{tabular}{p{3.0cm}p{2.0cm}p{2.0cm}p{2.0cm}p{1.8cm}p{1.8cm}}
		\hline
		Initial conditions  &  Normal stem cells ($\%$)  &  Normal cells ($\%$)   &   Tumor stem cells ($\%$) &    Tumor cells ($\%$) &  Time (months)$^{(a)}$\\
		\hline
		Case 1   &  2.35 &  11  & 2.62  & 81.77  & 1\\
		Case 2   &  5.2 &  4.7  & 6.4 & 81.87 & 14\\
		Case 3   &  0.46 &  0.15  & 13.19  & 84.51 & 26\\
		Case 4   &  3.95 &  5.04  & 3.65  & 85.36  & 63\\
		\hline
	\end{tabular}
	\label{tab:2}
	\begin{minipage}{14.5cm}
	$^{(a)}$ Time represents the duration required for the system to reach the dominant state (tumor cells $> 80\%$).
	\end{minipage}
\end{table*}

\begin{figure}[htbp]
	\centering
	\begin{center}
		\includegraphics[width=8cm]{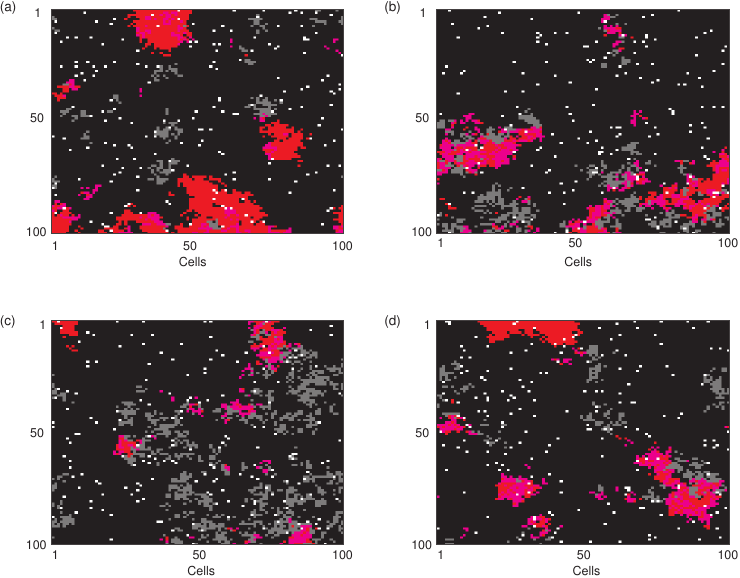}
	\end{center}
	\caption{\textbf{Spatial distribution of cell populations under different initial conditions.} \textbf{(a)} Case 1: Initialization with co-culture of normal and tumor stem cells ($m \in (0, 0.3)$). \textbf{(b)} Case 2: Initialization with sparse normal stem cells ($m \in (0, 0.3)$). \textbf{(c)} Case 3: Initialization from a stable healthy tissue state. Mutation rates for (a)-(c) were set to $(\alpha_1, \alpha_2) = (2.2\times 10^{-4}, 4.0\times 10^{-4})$. \textbf{(d)} Case 4: Initialization from a healthy tissue state with seeded tumor stem cells, but zero mutation rates. Results show representative spatial snapshots from a single stochastic realization. Note the similar terminal states despite vastly different time scales (see Table \ref{tab:2}).}
	\label{fig:7}
\end{figure}

Figure \ref{fig:7} displays the final spatial distributions for these scenarios. Although the pathways and timescales differed greatly, the endpoint patterns are remarkably similar, characterized by massive tumor occupancy and sparse remnants of normal tissue.

\begin{figure}[htbp]
	\centering
	\includegraphics[width=8cm]{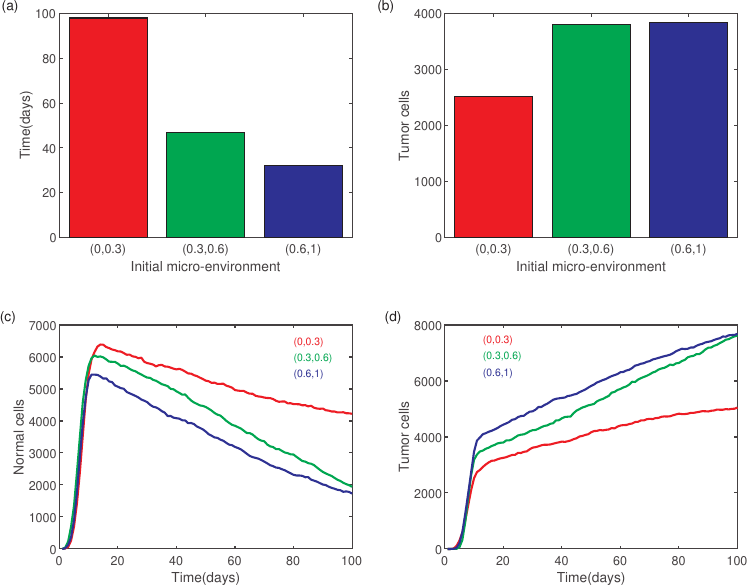}
	\caption{\textbf{Effects of initial microenvironmental conditions on tumor development.} \textbf{(a)} Time required for the tumor population to reach $5,000$ cells under different initial microenvironmental ranges. \textbf{(b)} Total tumor cell count at day $100$. \textbf{(c)} Temporal dynamics of the normal cell population. \textbf{(d)} Temporal dynamics of the tumor cell population. Data were obtained from a single representative stochastic realization.}
	\label{fig:8}
\end{figure}

To further decouple the role of the microenvironment from cellular density, we conducted simulations varying only the initial range of the microenvironmental value $m$ from $(0,  0.3)$, $(0.3,  0.6)$, and $(0.6, 1)$ while keeping cellular parameters constant. As shown in Figure \ref{fig:8}a, the time required for the tumor population to reach $5,000$ cells was inversely proportional to the initial $m$ value. In a hostile, high-$m$ environment, protection collapses, and tumors grow rapidly. Conversely, low-$m$ environments delay this process significantly. Figure \ref{fig:8}b confirms that by day $100$, tumor burden is markedly higher in initially deteriorated microenvironments. This pattern is consistently reflected in the population dynamics (Figures \ref{fig:8}c-d): as the initial microenvironment becomes less favorable (higher $m$), the normal cell population collapses more rapidly, while the tumor cell population exhibits accelerated exponential growth.

These findings highlight the critical duality of the tumor microenvironment: a favorable (low $m$) environment acts as a potent tumor-suppressive barrier that sustains homeostasis and delays malignancy---even in the presence of tumor seeds---whereas a deteriorated (high $m$) environment provides a permissive niche that catalyzes rapid neoplastic expansion. 

\subsection{Effects of microenvironmental changes on tumor cells growth}

\subsubsection{Microenvironmental intervention}

To examine the therapeutic potential of microenvironmental modulation, we simulated the administration of therapeutic agents designed to inhibit the tumor-induced microenvironmental transition, governed by the coefficient $k_{11}$. In the baseline (untreated) setting, $k_{11}=1$. We evaluated two therapeutic scenarios: a \textit{complete blockade} ($k_{11}=0$) and a \textit{partial inhibition} ($k_{11}=0.2$) of microenvironmental deterioration.

\begin{figure*}[htbp]
	\centering
	\begin{center}
		\includegraphics[width=13cm]{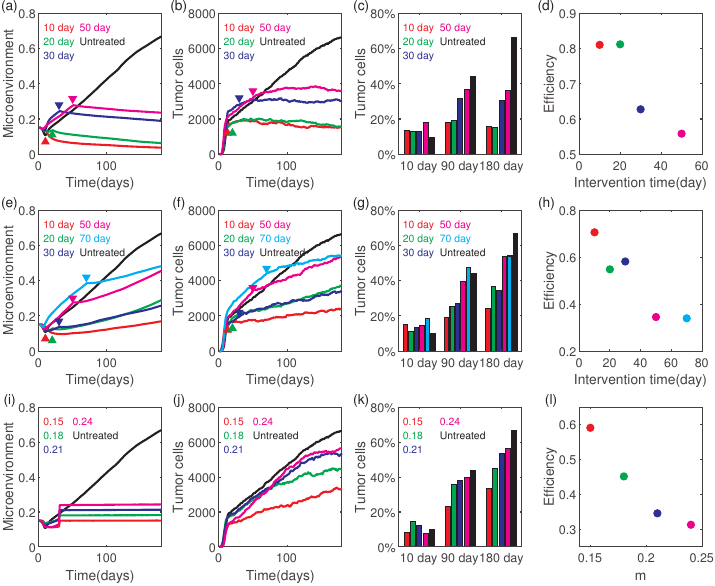}
	\end{center}
	\caption{\textbf{Impact of microenvironmental interventions on tumor cells growth.} Dynamics of average microenvironmental values under different therapeutic strategies: \textbf{(a)} complete blockade ($k_{11}=0$), \textbf{(e)} partial inhibition ($k_{11}=0.2$). \textbf{(i)} microenvironmental stabilized at specific fixed value $m$. Tumor cell population dynamics under the same conditions: \textbf{(b)} $k_{11}=0$, \textbf{(f)} $k_{11}=0.2$, \textbf{(j)} fixed $m$ values. Tumor cell fractions at specific developmental stages (Days $10$, $90$, $180$) for varied intervention timings: \textbf{(c)} $k_{11} = 0$, \textbf{(g)} $k_{11} = 0.2$, \textbf{(k)} fixed $m$ values. Dependence of therapy efficiency on intervention initiation time: \textbf{(d)} $k_{11} = 0$, \textbf{(h)} $k_{11} = 0.2$, and \textbf{(l)} on the fixed $m$ value. The efficiency is defined as the relative tumor cell reduction at day $180$ compared to the untreated condition. Results were obtained from a single representative realization for each strategy.}
	\label{fig:9}
\end{figure*}

We first simulated the complete blockade scenario ($k_{11}=0$), representing an ideal treatment that fully blocks the ability of tumor cells to remodel their surroundings. Interventions were initiated on days $10$, $20$, $30$, and $50$. In the absence of treatment, the microenvironment progressively deteriorated (Figure \ref{fig:9}a), driving sustained tumor expansion (Figure \ref{fig:9}b). In contrast, therapeutic intervention arrested this deterioration, maintaining the average microenvironmental value $m$ at a lower, homeostatic level (Figure \ref{fig:9}a). Consequently, tumor growth was significantly suppressed compared to the untreated case (Figure \ref{fig:9}b). 

Figure \ref{fig:9}c compares tumor cell fractions at days $10$, $90$, and $180$ across different intervention schedules. Early intervention shows smaller tumor cell fractions. To further quantify the dependency of therapeutic efficacy on timing, the relationship between efficacy and the intervention start day is shown in Figure \ref{fig:9}d. The results reveal a clear non-linear dependence: earlier treatment consistently yields a stronger inhibitory effect, resulting in an efficiency of $ 0.8$. In contrast, delaying intervention to day 30 has an efficiency dropping to $0.6$, and to $0.55$ when delaying to day 50. These findings are corroborated by the numerical data in Table \ref{tab:3}, which show that delaying treatment significantly compromises the suppression of tumor burden. This non-linear benefit of treatment efficiency suggests that TME-targeted therapies are most effective when applied before the microenvironment crosses the critical deterioration threshold.

\begin{table*}[t]
	\centering
	\caption{Tumor cell counts at different time points under complete blockade ($k_{11}=0$)}
	\begin{tabular}{p{4cm}p{2cm}p{2cm}p{2cm}p{2cm}p{2cm}}
		\hline
		Treatment Start Day    &  Day $10$  &  Day $20$ & Day $30$ &  Day $50$  &  Untreated \\
		\hline
		day $10$   & 1329   & 1301  &  1297 &  1818 & 970 \\
		day $90$  &  1803 &  1918 &   3143 &  3682 &  4397\\
		day $180$   &  1548  & 1536  & 3049  & 3616  & 6660 \\
		\hline
	\end{tabular}
	\label{tab:3}
\end{table*}

We next investigated a more clinically realistic scenario of partial inhibition ($k_{11}=0.2$), with treatments initiated on days $10$, $20$, $30$, $50$, and $70$. Similar to the complete blockade case, untreated simulations showed rapid microenvironmental deterioration (Figure \ref{fig:9}e) and unchecked tumor growth (Figure \ref{fig:9}f). Treatment provided partial restoration of the microenvironment, leading to a reduction in tumor growth rate. Consistent with the first scenario, earlier interventions resulted in substantially lower tumor burden, as illustrated in Figure \ref{fig:9}g-h. Table \ref{tab:4} confirms this trend; for instance, varying the starting day from $10$ to $70$ results in a significant difference in the final tumor cell count at day $180$.

\begin{table*}[t]
	\centering
	\caption{Tumor cell counts at different time points under partial inhibition ($k_{11}=0.2$)}
	\begin{tabular}{p{2cm}p{1.2cm}p{1.2cm}p{1.2cm}p{1.2cm}p{1.2cm}p{1.8cm}}
		\hline
		Treatment Start Day   &  Day $10$   &  Day $20$ &  Day $30$ &  Day $50$ &  Day  $70$   &  Untreated\\
		\hline
		day $10$  & 1537   & 1097 & 1329 & 1459  & 1858 &  970 \\
		day $90$  &  1917  &  2558 & 2701 & 3974  & 4762  &  4397\\
		day $180$  &  2402  &  3696  &  3421 & 5362 & 5407 & 6660\\
		\hline
	\end{tabular}
	\label{tab:4}
\end{table*}

We also evaluated a normalization strategy where the microenvironment was stabilized at fixed values ($m=0.15, 0.18, 0.21, 0.24$) following intervention on day $30$. Simulation results (Figure \ref{fig:9}i-l) demonstrated that maintaining a lower fixed value of $m$ significantly inhibited tumor growth. These results underscore the critical role of microenvironment homeostasis in limiting tumor progression.

\begin{figure*}[htbp]
	\centering
	\includegraphics[width=13cm]{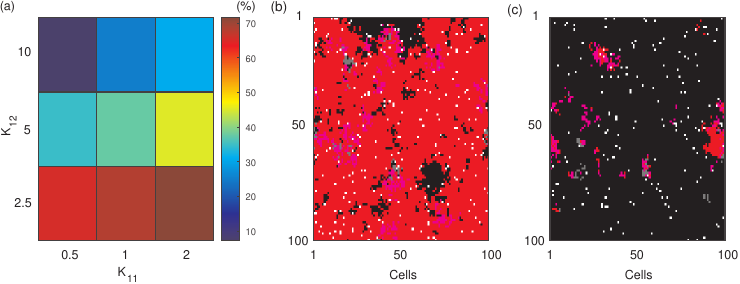}
	\caption{\textbf{Influence of microenvironmental transition kinetics on tumor burden.} \textbf{(a)} Proportion of tumor cells at day $100$ for various combinations of the maximum deterioration rate $k_{11}\in \{0.5, 1, 2\}$ and the saturation threshold $k_{12}\in \{2.5, 5, 10\}$. Results were averaged over $5$ independent simulation runs. \textbf{(b)} Spatial distribution of cells for the ''best-case'' scenario ($k_{11} = 0.5$, $k_{12} = 10$), showing minimal tumor burden due to high resistance to environmental changes. \textbf{(c)} Spatial distribution of cells for the ''worst-case'' scenario ($k_{11} = 2$, $k_{12} = 2.5$), showing maximal tumor efficacy driven by aggressive environmental remodeling. Results in (b) and (c) show representative spatial snapshots from a single stochastic realization.}
	\label{fig:10}
\end{figure*}

To gain mechanistic insight into how the kinertics of tumor-induced microenvironmental deterioation influence tumor evolution, we systematically explored two key parameters described in the Hill function: the maximum deterioration rate ($k_{11} \in \{0.5, 1, 2\}$) and the deterioration saturation threshold ($k_{12} \in \{2.5, 5, 10\}$). The tumor cell proportions on day $100$ are summarized in Figure \ref{fig:10}a. We observed that tumor burden is heavily dependent on the aggressiveness of this remodeling process. The lowest tumor burden occurred when the deterioration potential was minimized ($k_{11} = 0.5$) and the activation threshold was high ($k_{12} = 10$) (Figure \ref{fig:10}b). Conversely, the highest tumor burden occurred under conditions of rapid deterioration ($k_{11} = 2$) and high sensitivity to tumor signals (low threshold) ($k_{12} = 2.5$) (Figure \ref{fig:10}c). These results indicate that therapeutic strategies should aim to simultaneously limit the maximum rate of environmental deterioration (low $k_{11}$) and raise the threshold required for such changes (increase $k_{12}$), effectively desensitizing the microenvironment to tumor signals.

\subsubsection{Combined effects of mutation and microenvironment}

Having analyzed the effects of gene mutation and microenvironmental conditions separately \citep{2023Cancer,Marusyk:2010aa}, we next investigated their combined impact on tumor cell development. Simulations were performed by varying the initial microenvironmental value $m\in(0,1)$ (divided into $10$ intervals) and the mutation rate of normal stem cells $\alpha_{1}\in(10^{-6}, 10^{-1})$ (sampled at $11$ logarithmically spaced values), while keeping all other parameters fixed. Each simulation began with $20$ normal stem cells.

\begin{figure}[t]
	\centering
	\includegraphics[width=8cm]{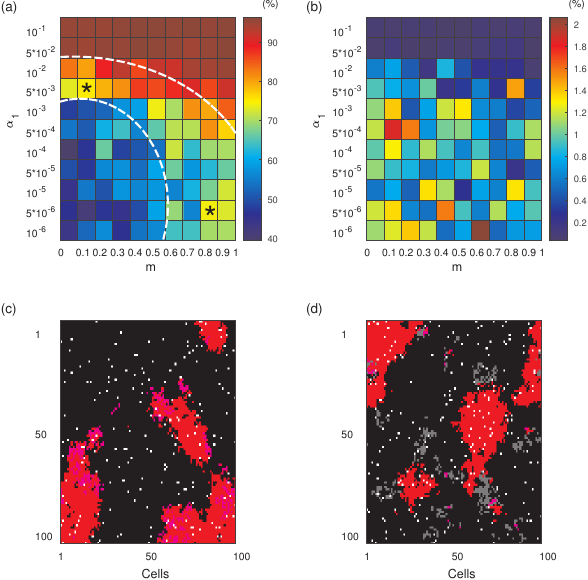}
	\caption{\textbf{Synergistic effects of initial microenvironment and gene mutation rates on tumor progression.} \textbf{(a)} Mean proportion of tumor cells on day $100$ under varying initial microenvironment values ($m$) and mutation rates ($\alpha_1$). \textbf{(b)} Standard deviation (SD) of tumor cell proportions on day $100$, illustrating the variability of simulation outcomes. High SD at low $\alpha_1$ indicates stochastic outcomes, while low SD at high $\alpha_1$ indicates deterministic tumor dominance. \textbf{(c)} Spatial distribution of cells driven by high mutation rate ($\alpha_1 = 5\times 10^{-3}$) in a favorable environment ($m \in (0.1, 0.2)$). \textbf{(d)} Spatial distribution of cells driven by a hostile environment ($m \in (0.8, 0.9)$) despiate a low mutation rate ($\alpha_1 = 5 \times 10^{-6}$). Data in (a) and (b) represent mean and SD derived from $n=5$ independent simulations. Panels (c) and (d) show representative spatial snapshots from a single stochastic realization.}
	\label{fig:11}
\end{figure}

Figure \ref{fig:11}a shows the mean tumor cell proportions on day $100$. Tumor burden exhibited a non-linear dependence on both parameters. In regions with low mutation rates $\alpha_1 < 10^{-3}$ and a favorable microenvironment ($m < 0.6$) (``Region 1''), tumor growth was minimal. However, a sharp phase transition occurred when $\alpha_1 \geq 10^{-2}$ (``Region 3''), indicating a mutation rate threshold beyond which tumor proliferation accelerates regardless of the initial environment. 

To quantitatively validate the distinct behavior observed across these regimes, we performed an independent Student's t-test ($n=5$). We defined the transition zone between Region 1 and Region 3 as ``Region 2''. The analysis revealed statistically significant differences between all adjacent regions: Region 1 vs. Region 2 ($p < 10^{-5}$), Region 2 vs. Region 3 ($p < 10^{-21}$), and Region 1 vs. Region 3 ($p < 10^{-40}$). These results confirm that the observed phase transitions represent robust shifts in system dynamics rather than stochastic fluctuations.

To further assess the stochasticity nature of these outcomes, we analyzed the standard deviation (SD) of the tumor cell proportions (Figure \ref{fig:11}b). In the transition zone ($m \in (0, 1)$ and $\alpha_1 \in (10^{-6}, 10^{-2})$),  the SD was relatively large, indicating high variability where tumor fate is sensitive to stochastic events. In contrast, when $\alpha_1 \geq 10^{-2}$, the SD became minimal. This suggests that high mutation rates drive the system into a stable, deterministic regime where aggressive tumor dominance is the inevitable outcome.

Figure \ref{fig:11}c and \ref{fig:11}d display spatial distributions for two distinct parameter combinations that yield similar total tumor counts: (c) a favorable environment ($m \in (0.1, 0.2)$) with high mutation rate ($\alpha_1 = 5\times10^{-3}$), and (d) a hostile environment ($m \in (0.8, 0.9)$) with low mutation rate ($\alpha_1 = 5\times10^{-6}$). Despite the different driving forces, the resulting spatial patterns are comparable, suggesting that intrinsic genetic instability and extrinsic environmental deterioration can act as interchangeable drivers of tumor progression.

These results collectively indicate that genetic mutations and microenvironmental deterioration function both independently and synergistically. Strategies that simultaneously improve microenvironmental conditions and reduce mutation rates are therefore essential for effective tumor control.

\section{Discussion}

Understanding the multiscale mechanisms that drive tumor cell emergence and progression is fundamental to developing effective cancer prevention and treatment strategies. In this study, we developed a two-dimensional cellular automaton model to simulate tumor evolution within a spatially heterogeneous microenvironment. The model distinguishes among several cell types---normal stem cells, normal non-stem cells, tumor stem cells, and tumor cells---allowing for detailed investigation of cell transitions and their interactions with the local microenvironment. By leveraging the discrete nature of CA modeling, we explicitly incorporated stochasticity in cell behavior, tracked spatial organization, and simulated individual-level variability in cellular dynamics.

Consistent with our model assumptions, our simulations demonstrate that the emergence of tumor cells is fundamentally dependent on nonzero gene mutation rates. When the mutation rate is zero, tumor cells fail to develop, confirming the essential role of genetic instability in oncogenesis. However, once the mutation rate becomes non-zero, tumor cells undergo rapid expansion, utilizing their competitive advantages in resource acquisition and spatial occupation. Under various combinations of mutation rates and initial microenvironmental conditions, tumor proliferation varies significantly. Notably, microenvironmental cues exert a profound influence on tumor progression: a benign or stable microenvironment can delay or suppress tumor expansion, while a hostile or disordered microenvironment significantly accelerates tumor growth. Furthermore, we observed that a moderate microenvironment is capable of retarding the expansion of tumors driven by genetic mutations. These findings are substantiated by our intervention simulations, which show that therapeutic strategies targeting the microenvironmental deterioration rate ($k_{11}$) or restoring the microenvironmental state ($m$) can effectively contain tumor progression.

By allowing individual cells to interact with and respond to both neighboring cells and environmental changes, the CA framework captures important features of tumor heterogeneity, including clonal diversity, spatial patterning, and nonlinear growth dynamics. Our results highlight the sensitivity of tumor growth outcomes to gene mutations and the dynamics of environmental remodeling. Specifically, we demonstrate that attenuating the tumor's capacity to induce environmental deterioration---either by reducing the maximum transition rate (low $k_{11}$) or by raising the saturation threshold required for this transition (high $k_{12}$)---can effectively inhibit tumor formation by preserving homeostatic conditions. Conversely, aggressive remodeling kinetics, characterized by a high transition rate ($k_{11}$) and low saturation threshold ($k_{12}$), significantly catalyze malignant expansion. Furthermore, both the initial microenvironmental state and the mutation rate of normal stem cells shape the eventual spatial distribution of the tumor, suggesting that early-stage environmental and genetic perturbations decisively influence tumorigenesis.

The model presented here offers a valuable tool for simulating tumor development in a computationally tractable yet biologically informative way. While continuum-based approaches (ODEs/PDEs) are powerful for describing population-level kinetics, our discrete framework offers distinct advantages for studying tumor microdynamics. Unlike well-mixed ODE models, the CA explicitly captures local spatial interactions and the formation of irregular tumor patterns. The model inherently accounts for the randomness of single-cell events (e.g., mutation, death), which is often smoothed out in continuum models but is critical for understanding rare evolutionary events. Moreover, it allows complex tissue-level phenomena (e.g., tumor dormancy versus outbreak) to spontaneously emerge from simple local rules governing cell-environment crosstalk. These advantages are well-suited for investigating how microscopic changes can give rise to macroscopic patterns of tumor progression and for providing insights that inform the development of predictive tools for cancer risk assessment and treatment planning, especially in the context of personalized medicine. Furthermore, we have made the source code publicly available to promote reuse and reproducibility, thereby contributing a flexible computational tool to the field of quantitative oncology.

Nevertheless, several limitations should be acknowledged. First, our implementation is currently restricted to a two-dimensional lattice. Extending the model to three dimensions would provide a more accurate depiction of tumor morphology and nutrient diffusion limits. Second, we simplified the microenvironment into a single scalar variable ($m$). Real physiological environments involve a complex interplay of immune cells, vasculature, extracellular matrix stiffness, and metabolic factors, which are not currently explicitly modeled. Third, we assumed a fixed set of behavioral rules for each cell type. Future interactions should incorporate dynamic phenotypic plasticity, such as epithelial-mesenchymal transition (EMT), to better simulate late-stage invasion and metastasis. Currently, the model also lacks distinct modules for complex biological feedback mechanisms, such as mechanical interactions or detailed signaling pathways. Finally, the simulation was conducted on a $100\times 100$ grid representing a small tissue patch. Hence, potential limitations may emerge from the spatial dimensionality and finite grid size of the simulations, particularly when scaling to whole-organ levels, which will require addressing computational efficiency challenges. Incorporating these biological features into future high-performance CA formulations will further enhance their realism and translational relevance.

Despite these limitations, the computational framework presented here exhibits a high degree of modularity and extensibility. The rule-based nature of the CA model allows distinct biological processes (e.g., cell division, mutation, resource consumption) to be treated as independent modules. This architectural flexibility serves as a robust platform for future research; additional layers of complexity---such as vascularization (angiogenesis), immune cell infiltration, or specific drug pharmacokinetic profiles---can be seamlessly integrated into the existing grid without disrupting the core evolutionary logic. Consequently, this model offers a scalable tool for testing comprehensive therapeutic strategies in a spatially resolved context.

In conclusion, this study highlights the utility of cellular automaton models in dissecting the coupling roles of genetic mutation and microenvironmental heterogeneity in tumor evolution. By capturing the stochastic, spatial, and nonlinear aspects of tumor growth, our model provides mechanistic insights into how favorable microenvironments serve as a barrier to tumorigenesis, while deteriorated environments catalyze malignant transformation. Future work will focus on integrating experimental data to calibrate parameters and applying the model to optimize combination therapies that simultaneously target tumor cells and their supportive niche.

\vspace{10pt} \noindent
{\bf Acknowledgements:}
This work was supported by the National Natural Science Foundation of China under grant No.12331018.

\noindent{\bf Availability:}
The source code of the cellular automata model is available at https://github.com/jinzhilei/Cellular\_Automata\_TME. Code implementation details are given in Appendix \ref{appa}.

\begin{appendix}
\section{Code Implementation Details}
\label{appa}

The computational model was implemented in \texttt{Python}. The source code is organized into modules to separate model definition, state updates, and numerical calculations. The key components of the code are described below:
\begin{enumerate}
\item Core Simulation Engine
\begin{itemize}
\item \texttt{main.py}: Serves as the main entry point for the simulation. It orchestrates the simulation loop, manages data input/output, and triggers the visualization of results.
\item \texttt{CellularAutomata.py}: Defines the spatial structure of the model. This module is responsible for constructing the cellular automaton (CA) grid (e.g., the hexagonal lattice) and establishing neighboring relations for each grid point to handle local interactions.
\item \texttt{StateUpdate.py}: Executes the iterative update process at each time step ($\Delta t$). It scans the grid and updates the status of each cell and grid point based on the stochastic outcomes derived from the model dynamics.
\end{itemize}
\item Biological Dynamics and Calculations
\begin{itemize}
\item \texttt{CellState.py}: Encapsulates the logic for various cellular behaviors and phenotypes. It defines the rules and conditions for cell proliferation, differentiation, migration, quiescence, and apoptosis (cell death) as described in the mathematical model.
\item \texttt{Calculation.py}: Contains the computational algorithms for state transitions. It explicitly implements the mathematical formulas (e.g., Hill functions for proliferation and fitness functions for death) to calculate each cell's transition probabilities based on its local microenvironment and internal state.
\end{itemize}
\item Initialization and Configuration
\begin{itemize}
\item \texttt{Initialization.py} series: These scripts generate the initial conditions for different simulation scenarios. They specify the initial spatial distribution of cell types (stem vs. non-stem, normal vs. tumor) and the initial microenvironmental landscape ($m$) corresponding to the specific Cases discussed in the Results section (e.g., healthy tissue vs. pre-seeded tumor).
\item \texttt{input\_individual} and \texttt{input\_public}: These files serve as configuration modules containing the model parameters. They store the numerical values for mutation rates ($\alpha$), proliferation rates ($\beta$), and microenvironmental interaction coefficients ($k_{11}, k_{12}$, etc.), allowing for easy parameter adjustment and sensitivity analysis.
\end{itemize}
\end{enumerate}
\end{appendix}

\bibliographystyle{elsarticle-harv}
\bibliography{HSC-2023}

\end{document}